\documentclass[pre,preprint,showpacs]{revtex4-1}
\usepackage{graphicx}
\usepackage{amsmath}
\usepackage{amssymb}
\usepackage{enumitem}

\begin{document}
\title{Interrupted coarsening in the zero-temperature kinetic Ising chain
driven by a periodic external field}
\author{Su Do Yi}
\affiliation{Department of Physics, Pukyong National University, Busan 608-737,
Korea}
\author{Seung Ki Baek}
\email{seungki@pknu.ac.kr}
\affiliation{Department of Physics, Pukyong National University, Busan 608-737,
Korea}

\begin{abstract}
If quenched to zero temperature, the one-dimensional Ising spin chain
undergoes coarsening, whereby the density of domain walls decays algebraically
in time. We show that this coarsening process can be
interrupted by exerting a rapidly oscillating periodic field with enough
strength to compete with the spin-spin interaction. By analyzing correlation
functions and the distribution of domain lengths both analytically and
numerically, we observe nontrivial correlation with more than one length scale
at the threshold field strength.
\end{abstract}


\maketitle

\section{Introduction}

One of the most important topics in statistical physics is the formation of
order. A classical nonequilibrium example is provided by
the one-dimensional (1D) Glauber-Ising model quenched to a zero temperature.
It approaches one of the ordered ground states by forming larger and larger
domains~\cite{bray}, and this coarsening process has been analyzed in full
detail (see, e.g., Ref.~\onlinecite{exact}). The relaxation toward
equilibrium is very slow: In the absence of an external field, the domain
walls perform annihilating random walks and the density accordingly decays as
$\rho \sim t^{-1/2}$ as time $t$ goes by~\cite{kinetic}.
A kinetically constrained version also exhibits glassy behavior with anomalously
slower coarsening~\cite{const,solich,ritort}, and such "aging" can even
cease to proceed under steady driving~\cite{fielding}.
One may ask if something similar can be achieved in the original
Glauber-Ising system by driving it with a suitable protocol, as has
been pointed out in Ref.~\onlinecite{bbk}.
Evidently, a constant external field does not work for that purpose
because the field will only accelerate the coarsening dynamics by breaking the
up-down symmetry. If the symmetry is concerned, an alternative
protocol would be an oscillating field with a short period.
This is of particular interest from the perspective of
interaction of light and matter in the high-frequency regime.
The problem becomes highly nontrivial, especially when the matter has
internal spatiotemporal correlations as a many-body system (see, e.g.,
Refs.~\onlinecite{leung,review,dsr} and references therein).
Due to its ubiquity and often dramatic consequences
as reported in Ref.~\onlinecite{kurchan},
the nonequilibrium caused by oscillatory driving still remains as an
active area of research to be explored further~\cite{verley,alt}.
If we consider the 1D Ising chain under an oscillating field,
one possible scenario is that the system has such a large
time scale that it simply overlooks the rapid oscillation so that
the field appears as a small perturbation around the ordered state.
Indeed, this has been numerically observed in the two-dimensional
kinetic Ising model subjected to an oscillating field (see, e.g.,
Ref.~\onlinecite{rikvold}).
On the other hand, it also seems plausible that the disordered state
can remain stable, although energetically unfavorable,
just as an inverted pendulum is stabilized by fast oscillatory
driving~\cite{huber,grebogi}. In this paper, we show that the latter
is the case when the field strength is greater than or equal to the
spin-spin coupling strength. In fact, from the dynamic rules
defined below, we can readily convince ourselves that
all the correlations are completely
destroyed if the field amplitude exceeds the spin-spin coupling strength,
which effectively corresponds to an infinite temperature.
If the field is weaker than the spin-spin interaction, on the other hand,
the up-down symmetry can be broken as in an ordered phase because
the system cannot escape from the absorbing states with all the spins
aligned in one direction.
Only when the internal and external energy scales are equally strong,
we observe finite nontrivial correlations and a stationary
density of domain walls. We will explain this point by calculating
correlation functions and the distribution of domain lengths both
analytically and numerically.

This paper is organized as follows: An explanation of our model
system is given in Sec.~\ref{sec:model}. Correlation functions
and the domain length distribution are analyzed in Sec.~\ref{sec:result}.
This is followed by a discussion of results and conclusions.

\section{Glauber-Ising dynamics}
\label{sec:model}

Let us consider a 1D Ising chain with size $L$ under a time-dependent external
field $H(t)$. The energy function is written as
\begin{equation}
\label{eq:e}
E =-\frac{J}{2}\sum_{i=1}^L S_i S_{i+1} - H(t)\sum_{i=1}^L S_i,
\end{equation}
where the spin variable $S_i$ can take either $+1$ or $-1$ and 
$\frac{J}{2} > 0$ is the coupling strength between neighboring spins.
We will impose a periodic boundary condition by setting $S_{L+1} = S_1$.
The time evolution of this system is assumed to obey the
zero-temperature Glauber dynamics~\cite{glauber}, which means that
every spin flips with the following rate:
\begin{equation}
\label{eq:w}
W_i =
\left\{
\begin{array}{lcl}
1           & \text{ if } & \Delta E_i < 0, \\
\frac{1}{2} & \text{ if } & \Delta E_i = 0, \\
0           & \text{ if } & \Delta E_i > 0,
\end{array}
\right.
\end{equation}
where $\Delta E_i$ is the energy difference due to a spin flip from
$S_i$ to $-S_i$. The external field $H(t)$ takes
a rectangular pulse shape between $+H_0$ and $-H_0$ with period $2T$,
where $H_0 > 0$ is a constant.
It is convenient to define $\tau \equiv (t \mod 2T)$ as a time index
within each period. Then, the external field is described as follows:
\begin{equation}
\label{eq:h}
H(t)=
\left\{
\begin{array}{lcl}
+H_0 & \text{ for } & 0 \leqslant \tau < T, \\
-H_0 & \text{ for } & T \leqslant \tau < 2T.
\end{array}
\right.
\end{equation}
As briefly mentioned above, we need to consider competition between the
spin-spin interaction and the external driving:
If $H_0 > J$, the field direction solely determines the dynamics, so that
the system is equivalent to a collection of non-interacting spins subjected to
the field.
If $H_0 < J$, on the other hand,
the field cannot flip a spin once it is surrounded by two other spins
in the same direction.
As a consequence, the density of domain walls keeps decreasing,
regardless of the field direction,
playing the role of the Lyapunov function in this dynamics.
This means that the steady states under periodic driving
must be the ordered ones for $H_0 < J$, and
the deterministic nature of the dynamics suggests that
the coarsening will not be slower than the field-free case.
One can indeed numerically check that the density of domain walls
decays as $\rho \sim t^{-1/2}$ when $T$ is small but with a smaller
prefactor than in the absence of $H(t)$.
For this reason, we can say that $H_0 = J$ is the most nontrivial point
due to the interplay between the field and the spin-spin interaction.
Henceforth, we will set $H_0 = J$ unless otherwise mentioned.

As is well known, the dynamics can also be analyzed in terms of
domain walls. We will briefly review three basic processes
of the domain-wall dynamics, i.e., pair creation, pair annihilation,
and propagation, assuming that $H(t)= +H_0$.
First, two domain walls are created inside a down-spin domain
$\{\cdots \downarrow\downarrow\downarrow \cdots\}$ when the field flips
the spin in the middle with rate $\frac{1}{2}$, which results in
$\{\cdots \downarrow\uparrow\downarrow \cdots\}$.
Second, the pair-annihilation process is possible in two different ways, i.e.,
$\{\cdots \uparrow\downarrow\uparrow \cdots\} \Longrightarrow
\{\cdots \uparrow\uparrow\uparrow \cdots\}$ with rate $1$ or
$\{\cdots \downarrow\uparrow\downarrow \cdots\}
\Longrightarrow \{\cdots\downarrow\downarrow\downarrow \cdots\}$
with rate $\frac{1}{2}$.
Last, a domain wall propagates when a spin flips at a domain boundary,
e.g., $\{\cdots \downarrow\downarrow\uparrow \cdots\} \Longrightarrow
\{\cdots \downarrow\uparrow\uparrow \cdots\}$ with rate $1$.
In the spin language, all these processes tend to align spins
along the field direction. Therefore,
few domain walls exist if the field has been applied for
a sufficiently long period. One of our primary interests is
how the density of domain walls varies in time when the time-dependent field
in Eq.~(\ref{eq:h}) drives the system.

\begin{figure}
\center\includegraphics[width=0.70\textwidth]{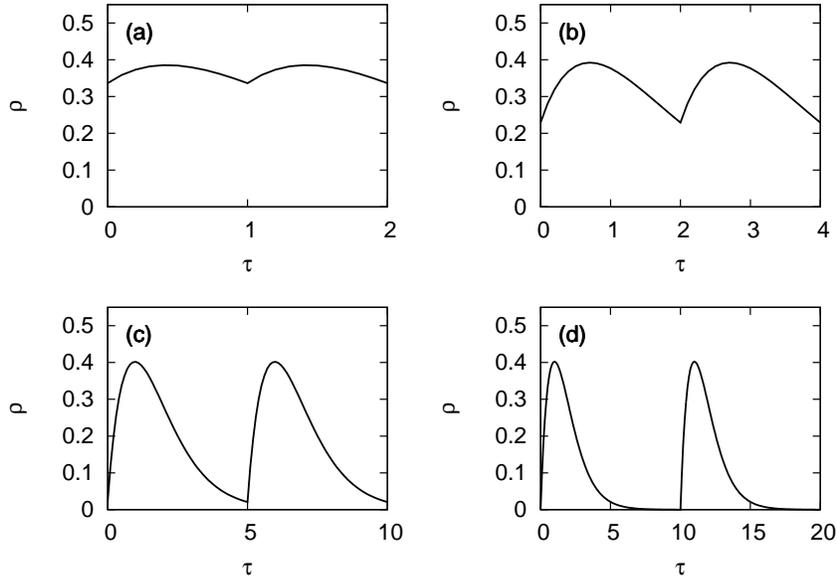}
\caption{Density of domain walls in the zero-temperature Ising chain of length
$L=10$, entrained by the field in Eq.~(\ref{eq:h}) with
(a) $T=1$, (b) $T=2$, (c) $T=5$, and (d) $T=10$.
Each panel shows numerically exact results from
Eq.~(\ref{eq:eigen}) and Monte Carlo results averaged over $10^5$ periods,
although they are indistinguishable in this plot.}
\label{fig:mc}
\end{figure}

\begin{figure}
\center\includegraphics[width=0.70\textwidth]{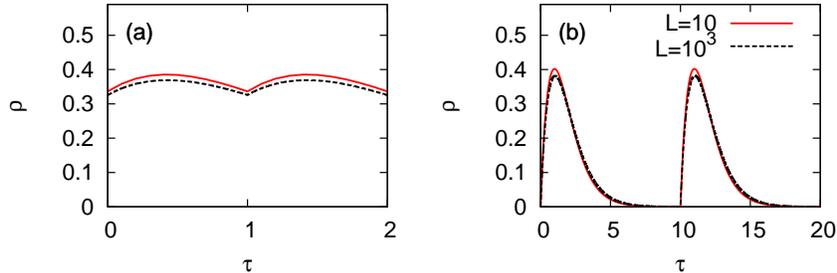}
\caption{(Color online) Size dependence in the density of domain walls for
(a) $T=1$ and (b) $T=10$, obtained by using Monte Carlo calculations,
where the solid (dotted) lines represent $L=10$ ($L=10^3$).}
\label{fig:size}
\end{figure}

We can formally describe the Glauber-Ising dynamics
by using the transition-matrix formulation because it is Markovian.
The Ising chain in Eq.~(\ref{eq:e}) has $N=2^L$ microstates.
Indexing the microstates by $\alpha = 1,\ldots,N$,
we define $p_\alpha(t)$ as the probability to find the system in state
$\alpha$ at time $t$. The probability distribution can then be denoted as
$\mathbf{p}(t) \equiv \left\{p_1(t), p_2(t), ... , p_N(t)\right\}$
with a constraint for the conservation of total probability,
$\sum_\alpha p_\alpha(t) = 1$.
One can readily calculate any single-time observable from $\mathbf{p}(t)$
in principle, including the average domain wall density.
The zero-temperature Glauber rates in Eq.~(\ref{eq:w}) define an $N \times N$
transition matrix $\mathbf{M}(t)$ that governs the evolution of $\mathbf{p}(t)$
in the following way:
\begin{equation}
\mathbf{p}(t+\Delta t) = \mathbf{M}(t)  \mathbf{p}(t),
\end{equation}
where $\Delta t$ means a time scale for flipping a single spin.
It is reasonable to suppose that every spin has a chance to flip during one
time step on average, which means that $\Delta t$ should be proportional to
$L^{-1}$.
The rates are dependent on the external field, so
we can distinguish the rates under $H(t) = +H_0$ from those under $-H_0$.
It implies that we have to work with two transition matrices:
\begin{equation}
\mathbf{M}(t)=
\left\{
\begin{array}{lcl}
\mathbf{M}^+ & \text{ if } & H(t)=+H_0, \\
\mathbf{M}^- & \text{ if } & H(t)=-H_0,
\end{array}
\right.
\end{equation}
which are actually related by a simple coordinate transformation~\cite{alt}.
After one period, therefore, the probability distribution at time $t=0$
evolves to $\mathbf{p}(t=2T) = \mathbf{M}_T \mathbf{p}(t=0)$
with $\mathbf{M}_T \equiv [(\mathbf{M}^-)^L]^T
[(\mathbf{M}^+)^L]^T$.
When the system has been entrained by the driving, it should be
found statistically identical at time $t$ and $t+2T$. This can be regarded
as a nonequilibrium steady state in a stroboscopic sense. For
example, we may observe the system at the beginning of every period, i.e.,
at $\tau = 0$ and denote the resulting steady state as
$\mathbf{p}_\infty (\tau=0)$.
It is obtained by solving the following equation:
\begin{equation}
\label{eq:eigen}
\mathbf{p}_\infty (\tau=0)= \mathbf{M}_T\mathbf{p}_\infty (\tau=0),
\end{equation}
and the existence of such an eigenvector is guaranteed because
both the $\mathbf{M}^+$ and the $\mathbf{M}^-$ are stochastic.
The steady-state distribution for general $\tau$ is also obtained in
a straightforward way.
In practice, Eq.~(\ref{eq:eigen}) can be solved only for
$L \lesssim O(10)$ because the size of $\mathbf{M}$ grows as
an exponential function of $L$.
From a computational point of view,
it is often more efficient to sample configurations by using a Monte Carlo
method.
Figure~\ref{fig:mc} demonstrates that the Monte Carlo sampling precisely
reproduces the result from Eq.~(\ref{eq:eigen}). Our Monte Carlo result also
shows that the transition-matrix calculation for $L=10$ is quite accurate in
estimating the average density of domain walls in a larger system
(Fig.~\ref{fig:size}).
It implies the following:
Suppose that we randomly take ten consecutive spins in a large system many times
and count the frequency of an arbitrary spin configuration $i$. Our observation
suggests that it will be more or less similar to $p_i$ obtained from the
transition-matrix calculation, and it is supported by Monte Carlo calculations
(not shown). If a large system can be approximated as a collection
of small ones of $L \sim O(10)$, it is because
the characteristic length scale is shorter than $O(10)$. In other words,
this observation suggests weakness of the interaction between domain walls.
This remark will also be supported by other observations below.

Another important question in this context is whether a dynamic phase
transition (DPT) occurs as the half period $T$ is varied. For example,
for dimensions higher than one, the Glauber-Ising model undergoes a
symmetry-breaking DPT at a sufficiently low temperature as $T$
decreases~\cite{rikvold1998,dpt}.
Such a DPT is explained by the competition between
internal and external time scales for relaxation and driving, respectively.
However, such a DPT seems unlikely in our Ising chain, although the temperature
is zero: One clue in Figs.~\ref{fig:mc} and \ref{fig:size}
is that the response to $+H_0$ ($ 0 \leqslant \tau < T$) is indistinguishable
from the one to $-H_0$~($ T \leqslant \tau < 2T$) for any value of $T$.
The magnetization $m= L^{-1} \sum_i S_i$ also oscillates around zero with
preserving the up-down symmetry for any $T$ (not shown). We will present a more
quantitative argument for the absence of a DPT by using correlation functions,
which we introduce below.

\section{Results}
\label{sec:result}

\subsection{Correlation functions}

We begin by considering slow driving, e.g., as in Fig.~\ref{fig:mc}(d).
One can easily understand the behavior of the density of domain walls:
At $\tau = 0$, for example, the field abruptly changes
from $-H_0$ to $+H_0$, whereas most of the spins are pointing downward.
The density of domain walls thus increases when $\tau$ is small.
As $\tau$ grows further, however, it is followed by a downturn in the density
because almost all the spins are aligned in the field direction.
Then, the field changes to $-H_0$ again, and all the processes of
creation and annihilation of domain walls are repeated anew.
We will put this description on a more quantitative ground by
considering correlation functions, and then move on to the case of fast
driving.

\begin{figure}
\center\includegraphics[width=0.70\textwidth]{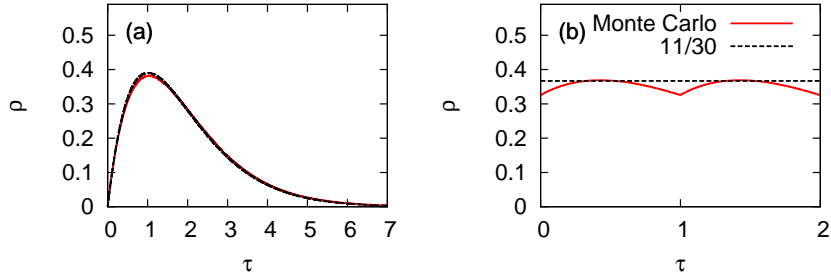}
\caption{(Color online)
Density of domain walls as a function of $\tau$.
(a) The solid lines are obtained by using Monte Carlo simulations for $L=10^3$
and $T=15$, whereas the dotted lines are obtained by numerically integrating
Eq.~(\ref{eq:eqmotion}).
(b) Monte Carlo results for $L=10^3$ and $T=1$ (solid line)
remain close to the value $\frac{11}{30}$ estimated from
Eq.~(\ref{eq:eqmotion}) in the limit of $T \rightarrow 0$ (dotted line).}
\label{fig:dw}
\end{figure}

Let us recap the time evolution of an individual spin $i$ during $\Delta t$ as
follows:
\begin{equation}
S_i(t+\Delta t) =
\left\{
\begin{array}{lcl}
 S_i(t) & \text{ with probability } & 1-W_i \Delta t, \\
-S_i(t) & \text{ with probability } & W_i \Delta t,
\end{array}
\right.
\end{equation}
where $W_i$ is given in Eq.~(\ref{eq:w}).
In the limit of $\Delta t \rightarrow 0$, the time derivative of
magnetization and that of the two-point correlation function can be written as
\begin{equation}
\label{eq:mag}
\frac{d\langle S_i\rangle}{dt}=- 2\left\langle S_i W_i \right\rangle,
\end{equation}
and
\begin{equation}
\label{eq:corr}
\frac{d\langle S_i S_{i+r}\rangle}{dt}=- 2\left\langle S_i S_{i+r}
(W_i +W_{i+r}) \right\rangle,
\end{equation}
respectively, where $\langle \cdots \rangle$ means the average over
configurations.
We now suppose that the system experiences $+H_0$.
Enumerating all the possible spin triplets, we can summarize the
Glauber transition rates in Eq.~(\ref{eq:w}) as follows:
\begin{equation}
\label{eq:rate}
W_i=\frac{1}{2} \left[ g_i + (1-g_i)(1-S_i) \right],
\end{equation}
with $g_i \equiv \frac{1}{4} (1-S_{i-1})(1-S_{i+1})$.
By substituting Eq.~(\ref{eq:rate}) with Eqs.~(\ref{eq:mag}) and~(\ref{eq:corr}),
we find that
\begin{subequations}
\label{eq:correlation}
\begin{align}
\frac{dm(t)}{dt}   &= \frac{1}{4}\left(3-2m-C_2\right),\label{eq:m}\\
\frac{dC_r(t)}{dt} &= \frac{1}{2}\left(  3m-4C_r +
C_{r-1}+C_{r+1}-C_{r-1,2}\right),
\end{align}
\end{subequations}
where $m \equiv \langle S_i \rangle$,
$C_r \equiv \langle S_i S_{i+r} \rangle$, and
$C_{l,r} \equiv \langle S_{i-l} S_i S_{i+r} \rangle = C_{r,l}$.
Note that we have
assumed invariance under translation and reflection in the correlation
functions.
We could also write down the evolution of the three-point correlation
functions, but it is already obvious that the equations will not be closed.
To proceed, we need to truncate the endless sequence of equations.
Our minimalist description is neglecting correlation over a distance
greater than two, so it reads as
\begin{subequations}
\label{eq:eqmotion}
\begin{align}
\frac{dm(t)}{dt}   &=\frac{1}{4}\left(3-2m- C_2\right),\label{eq:m0}\\
\frac{dC_1(t)}{dt} &=\frac{1}{2}\left(3 m-4C_1 + 1+C_2-m\right),\label{eq:c1}\\
\frac{dC_2(t)}{dt} &=\frac{1}{2}\left(3 m-4C_2 +   C_1 \right),\label{eq:c2}
\end{align}
\end{subequations}
where we have included the evolution of $C_2$, which appears in
Eq.~(\ref{eq:m}).
This description is minimalist in the following sense:
Suppose that $C_2$ is also neglected. Considering Eq.~(\ref{eq:m0}),
we see that this makes
the evolution of $m$ independent of other correlation functions. Unregulated
by higher-order correlations, it has a fixed point at $m=\frac{3}{2}$, which is
unphysical. We thus conclude that we need to take into account $C_2$ at least.
Note that our simplified dynamics still admits a fully ordered state with
$m = C_1 = C_2 = 1$ as a stationary solution for the static field $H(t) = +H_0$.
Calculating the density of domains walls,
$\rho(t) = \frac{1}{2} \left[ 1-C_1(t) \right]$,
we find a striking agreement between Monte Carlo results
and Eq.~(\ref{eq:eqmotion}), numerically integrated from
an initial condition with $m = -1$ and $C_1 = C_2 = 1$
[Fig.~\ref{fig:dw}(a)]. This agreement is also consistent with the remark
in the previous section that the correlation length is not greater than $O(10)$.

Having checked our description for slow driving,
we may now consider the opposite limit of $T \rightarrow 0$.
We assume that this limit restores the up-down symmetry
so that $m$ is negligible in Eqs.~(\ref{eq:c1}) and (\ref{eq:c2})
in an average sense.
This assumption is supported by the following argument:
If any remnant magnetization $m \neq 0$ exists, it means that
the system is unable to respond to such a rapid
field modulation. According to this idea, $C_2$ would not change appreciably
upon the field reversal, either.
When $H(t) = -H_0$, Eq.~(\ref{eq:m0}) takes a slightly different form:
\begin{equation}
\label{eq:negative}
\frac{dm(t)}{dt}=\frac{1}{4}\left(-3-2m+ C_2\right),
\end{equation}
where the right-hand side is written in terms of the same correlation
functions based on the "freezing" scenario above.
The summation of Eqs.~(\ref{eq:m0}) and~(\ref{eq:negative}) expresses
the total change in $m$ during one period, which must vanish in a steady state.
This immediately leads to the conclusion that $m=0$.
In this way, we can argue that a symmetry-breaking DPT, as a result
of the competition between relaxation and driving time scales, should be
absent in our system. Put differently,
the relaxational time scale does not grow longer than the one for driving,
and this is consistent with our observation of short correlation lengths.
As a side remark, we add that the statement of vanishing $m$ contains a subtle
point: If $m$ was strictly zero all the time, it would imply $\frac{dm}{dt}
\neq 0$ in Eq.~(\ref{eq:m0}) or (\ref{eq:negative}),
which is self-contradictory. A correct explanation is rather that
$m$ will keep changing around zero with a small magnitude.
In addition to $m=0$ in an average sense, the steady-state condition requires
that both $\frac{dC_1}{dt}$ and $\frac{dC_2}{dt}$ must also vanish.
Solving the set of linear equations resulting from Eqs.~(\ref{eq:c1}) and
(\ref{eq:c2}), we estimate the stationary density of domain walls as
$\rho = \frac{1}{2} (1-C_1) = \frac{11}{30}$.
This calculation agrees well with our Monte Carlo
result, confirming the existence of domain walls in the presence
of a fast switching field [Fig.~\ref{fig:dw}(b)].

\subsection{Domain statistics}

So far, we have focused on the lowest-order ones among the
infinite hierarchy of correlation functions, and this turns out to be
enough to describe certain average quantities, such as the density of domain
walls. Now, let us proceed to the detailed statistics of domains to gain
more information.
We begin by considering how domains evolve in time when the field is taken
to be $+H_0$. Let $P_n$ denote the density of down-spin domains of length $n$
so that $\sum_n n P_n$ is equal to the fraction of down spins.
We have four mechanisms that affect $P_n$.

\begin{enumerate}[label=(\arabic*)]
\item
A domain of length $n$ disappears when any of its down spins
flips upward. According to Eq.~(\ref{eq:w}), two spins at the boundary flip
with rate $1$, whereas the rate is reduced to $\frac{1}{2}$ for the other
$(n-2)$ spins in the bulk. Therefore, the total rate of loss amounts to
$1\times 2 + \frac{1}{2} \times (n-2) = \left(\frac{n}{2} + 1\right)$,
multiplied by $P_n$, for $n \ge 2$.
Note that this formula does not cover the case of a single-spin domain,
which disappears via $\{\cdots \uparrow\downarrow\uparrow \cdots\}
\Longrightarrow
\{\cdots \uparrow\uparrow\uparrow \cdots\}$ with rate $1$.
\item
The density $P_n$ increases when a domain of length $n+1$ shrinks by one at the
boundary. The contribution is counted as $2P_{n+1}$ because of the
two boundary spins.
\item
We can increase $P_n$ by dividing a domain of length $l \ge n+2$
into two pieces in such a way that
\begin{equation}
\{ \cdots \uparrow \underbrace{\downarrow \cdots \downarrow \downarrow
\downarrow \cdots \downarrow}_{l} \uparrow \cdots \}
\Longrightarrow
\left\{
\begin{array}{lcl}
\{ \cdots \uparrow \underbrace{\downarrow \cdots \downarrow}_{n} \uparrow
\underbrace{\downarrow \cdots \downarrow}_{l-n-1} \uparrow \cdots \}
& \text{ with rate } & \frac{1}{2},\\
\{ \cdots \uparrow \underbrace{\downarrow \cdots \downarrow}_{l-n-1} \uparrow
\underbrace{\downarrow \cdots \downarrow}_{n} \uparrow \cdots \}
& \text{ with rate } & \frac{1}{2}.
\end{array}
\right.
\end{equation}
If $n \neq l-n-1$,
this has two different possibilities, each with rate $\frac{1}{2}$, so
the contribution to $P_n$ from the domain of length $l$ is equal to $P_l$.
Even if $n = l-n-1$, the contribution is still $P_l$ because
the division creates two domains of length $n$ with rate $\frac{1}{2}$.
In total, this third mechanism contributes $\sum_{l=n+2}^{\infty} P_l$ to
$P_n$.
\item
The last mechanism is to merge a domain of size $l \le n-2$ and another
with size $n-l-1$ to create a domain of size $n$. We can visualize it as
\begin{equation}
\{ \cdots \uparrow \underbrace{\downarrow \cdots \downarrow}_{l} \uparrow
\underbrace{\downarrow \cdots \downarrow}_{n-l-1} \uparrow \cdots \}
\Longrightarrow
\{ \cdots \uparrow \underbrace{\downarrow \cdots \downarrow \downarrow
\downarrow \cdots \downarrow}_{n} \uparrow \cdots \} \text{ with rate }
\frac{1}{2}.
\end{equation}
To evaluate the probability of this event, we need to know
the probability of the configuration on the left-hand side.
The independent-interval approximation (IIA) suggests that the lengths can
be regarded as totally uncorrelated so that the probability
can be expressed as $P_{l} P_{n-l-1}$~\cite{iia}. The total contribution
of this mechanism is thus approximately written as
$\frac{1}{2} \sum_{l=1}^{n-2} P_l P_{n-l-1}$.
\end{enumerate}

Gathering all these terms, we arrive at
\begin{subequations}
\label{eq:dist}
\begin{align}
\frac{dP_1}{dt} &=-P_1 + 2P_2 + \sum_{l=3}^{\infty}P_l,\\
\frac{dP_n}{dt} &\simeq-\left(1+\frac{n}{2}\right)P_n + 2 P_{n+1} +
\sum_{l=n+2}^{\infty}P_l + \frac{1}{2}\sum_{l=1}^{n-2}P_l P_{n-l-1}
\text{ for }n \geq 2.
\end{align}
\end{subequations}

The next step is to consider the dynamics of up-spin domains with keeping
the same field direction. Similar to $P_n$, we define $Q_n$ as the density
of up-spin domains of length $n$. We have five mechanisms to affect $Q_n$.
\begin{enumerate}[label=(\roman*)]
\item In the first mechanism, a domain of a single up spin evaporates via
$\{\cdots \downarrow\uparrow\downarrow \cdots\} \Longrightarrow
\{\cdots \downarrow\downarrow\downarrow \cdots\}$ with rate $\frac{1}{2}$.
This takes place only for $n=1$.
\item Again, this second mechanism applies only to $n=1$. A domain of length
$1$ can be created via
$\{\cdots \downarrow\downarrow\downarrow \cdots\} \Longrightarrow
\{\cdots \downarrow\uparrow\downarrow \cdots\}$ with rate $\frac{1}{2}$.
For this to happen, we have to pick up a down spin surrounded by two other
down spins. For a down-spin domain of size $l$, we have $l-2$ such spins.
Therefore, we compute this contribution as $\frac{1}{2}
\sum_{l=3}^\infty (l-2) P_l$. Note that the dynamics of $Q_n$ is coupled
to that of $P_n$.
\item The third mechanism describes a loss due to the growth from length $n$.
The domain can grow to the left or right, each with rate $1$,
so the contribution becomes $-2Q_n$.
\item A domain of length $n$ can be gained from the growth process as well,
when a domain of length
$n-1$ expands to $n$ by flipping a spin upward at the boundary with rate $1$.
However, we cannot simply write it as $2Q_{n-1}$ because the spin flip
may merge this domain with another. For example, if we look at the left
boundary, the following process creates a domain of length $n$:
\begin{equation}
\{ \cdots \Downarrow \downarrow
\underbrace{\uparrow \cdots \uparrow}_{n-1} \downarrow \cdots \}
\Longrightarrow
\{ \cdots \Downarrow
\underbrace{\uparrow \uparrow \cdots \uparrow}_{n} \downarrow \cdots \},
\end{equation}
whereas the following does not:
\begin{equation}
\{ \cdots \Uparrow \downarrow
\underbrace{\uparrow \cdots \uparrow}_{n-1} \downarrow \cdots \}
\Longrightarrow
\{ \cdots \Uparrow
\underbrace{\uparrow \uparrow \cdots \uparrow}_{n} \downarrow \cdots \}.
\end{equation}
In short, it depends on the direction of the spin drawn as a double
arrow on the leftmost side.
In a similar spirit to the IIA, we assume a well-defined
probability $\Phi$ for the spin to point downward so that the contribution
becomes $2Q_{n-1}\Phi$.
\item The last mechanism is to merge two up-spin domains, one with size $l$
and the other with size $n-l-1$ as follows:
\begin{equation}
\{ \cdots \downarrow \underbrace{\uparrow \cdots \uparrow}_{l} \downarrow
\underbrace{\uparrow \cdots \uparrow}_{n-l-1} \downarrow \cdots \}
\Longrightarrow
\{ \cdots \downarrow \underbrace{\uparrow \cdots \uparrow \uparrow \uparrow
\cdots \uparrow}_{n} \downarrow \cdots \} \text{ with rate } 1.
\end{equation}
As before, we resort to the IIA to estimate the contribution as
$\sum_{l=1}^{n-2} Q_l Q_{n-l-1}$.
\end{enumerate}
To sum up, we have derived equations for $Q_n$ as
\begin{subequations}
\label{eq:dist2}
\begin{align}
\frac{dQ_1}{dt} &=-\frac{1}{2}Q_1 - 2 Q_1 +\frac{1}{2}\sum_{l=3}^{\infty}
(l-2) P_l,\\
\frac{dQ_n}{dt} &\simeq-2Q_n+2Q_{n-1} \Phi + \sum_{l=1}^{n-2}Q_l Q_{n-l-1}
\text{ for }n \geq 2.
\end{align}
\end{subequations}
Even if $H(t) = -H_0$, we can derive essentially the same as
Eqs.~(\ref{eq:dist}) and~(\ref{eq:dist2}), provided that
the variable $Q_n$ indicates domains in the direction of the field,
whereas $P_n$ does in the opposite direction.

\begin{figure}
\center\includegraphics[width=0.70\textwidth]{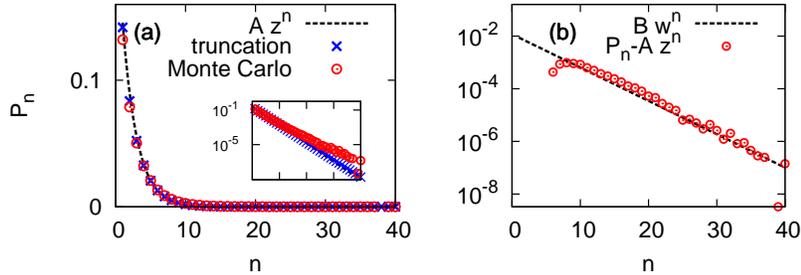}
\caption{(Color online)
Domain size distribution.
(a) The dotted line shows the trial solution $P_n= A z^n$ with
$A \simeq 0.236\,629$
and $z \simeq0.615\,633$ that approximately solves Eq.~(\ref{eq:dist3}).
The crosses represent a numerically exact solution of Eq.~(\ref{eq:dist3})
truncated at $n=50$.
The circles are Monte Carlo results for the Ising chain of length $L=10^4$.
The inset shows the same data in a semi-logarithmic plot.
(b) Correction from the simple exponential form [Eq.~(\ref{eq:correction})].
We estimate $B \simeq 0.01$ and $w \simeq 0.75$ by fitting the data on a
logarithmic scale.
}
\label{fig:ds}
\end{figure}

Suppose that $T$ is so short that the down-spin domains are
effectively subjected to both Eqs.~(\ref{eq:dist}) and~(\ref{eq:dist2}).
The steady-state condition implies that $\frac{dP_n}{dt} +
\frac{dQ_n}{dt} = 0$ for every $n \ge 1$. As the up-down symmetry is
restored, we may also equate every $Q_n$ with $P_n$ with setting $\Phi =
\frac{1}{2}$. We finally end up with the following set of equations:
\begin{subequations}
\label{eq:dist3}
\begin{align}
0 &=-\frac{7}{2} P_1 + 2P_2
+\sum_{l=3}^{\infty} \frac{l}{2} P_l,\label{eq:z}\\
0 &= P_{n-1} -\left(3+\frac{n}{2}\right)P_n + 2 P_{n+1} +
\sum_{l=n+2}^{\infty}P_l
+ \frac{3}{2}\sum_{l=1}^{n-2}P_l P_{n-l-1}
\text{ for }n \geq 2.\label{eq:a}
\end{align}
\end{subequations}
Our trial solution is an exponential distribution, i.e., $P_n = A z^n$ with
positive constants $A$ and $z<1$. Substituting this into Eq.~(\ref{eq:z}),
we obtain $z \simeq 0.615\,633$. It is worth noting that
$z$ would be equal to $\frac{1}{2}$ if all the correlations were destroyed
as in the infinite-temperature limit.
Although this trial solution does not exactly solve Eq.~(\ref{eq:a}),
we can estimate the amplitude $A \simeq 0.236\,629$ by
taking $n \rightarrow \infty$.
As a cross-check, we truncate Eq.~(\ref{eq:dist3}) by
setting $P_n = 0$ for $n>50$, and solve the $50$ coupled equations for
$P_1, \ldots, P_{50}$ simultaneously.
It confirms the validity of our trial solution even for small values of $n$
as shown in Fig.~\ref{fig:ds}(a). Of course, we have to ask ourselves whether
Eq.~(\ref{eq:dist3}), involved with several uncontrolled approximations,
correctly describes the domain dynamics. This is checked
by simulating an Ising chain of length $L=10^4$ to sample the domain length
distribution. As depicted in Fig.~\ref{fig:ds}(a), the result shows that
Eq.~(\ref{eq:dist3}) works qualitatively but tends to underestimate $P_n$
when $n$ is large.
The correction from $A z^n$ reveals another length scale in the following form:
\begin{equation}
\label{eq:correction}
P_n - Az^n \simeq B w^n,
\end{equation}
with $B \simeq 0.01$ and $w \simeq 0.75$ [Fig.~\ref{fig:ds}(b)].
The second length scale corresponds to roughly four lattice spacings, about
twice larger than the first one, but its the origin is not fully understood yet.

\section{Summary}

To summarize, we have considered the zero-temperature Glauber dynamics
in the 1D Ising chain driven by rectangular pulses of period $2T$ and strength
equal to $J$.
We have argued that the driving interrupts the coarsening so that
the density of domain walls converges to a nonzero stationary value
$\simeq \frac{1}{3}$ in the limit of fast driving.  We have also calculated
the steady-state distribution of domain lengths in the same limit
by using the IIA, and the result indicates the existence of finite nontrivial
correlation. Moreover, our Monte Carlo calculation shows that the actual density
is higher than expected from simple exponential decay, revealing the existence
of the second length scale, about twice larger than the first one.

\acknowledgments
We gratefully acknowledge discussions with J.-L. Barrat.
S.D.Y. was supported by Basic Science Research
Program through the National Research Foundation of
Korea funded by the Ministry of Education (Grant No. NRF-2014R1A6A3A01059435).
S.K.B. was supported by Basic Science Research Program through the
National Research Foundation of Korea funded by the Ministry of Science,
ICT and Future Planning (Grant No. NRF-2014R1A1A1003304 and No. NRF-2014K2A7A1044368),
and the program was coupled to Grant No. ERC-AdG-291073-Glassdef funded by the European
Research Council.

%
\end{document}